\documentclass[
english,aps,prb,amsmath,amssymb,twocolumn,letterpaper,citeautoscript,superscriptaddress,longbibliography]{revtex4-2}
\usepackage{svg}
\usepackage{graphicx}
\usepackage{dcolumn}
\usepackage{bm}

\usepackage{tikz}
\usepackage{hyperref}

\hypersetup{linkcolor=blue,urlcolor=blue,colorlinks,citecolor=blue}

\begin{document}

\title{Fractional Impurity Entropy from Potential Scattering in a Free Electron Gas}

\author{Syeda Neha Zaidi}

\affiliation{Department of Physics and Astronomy, Purdue University, West Lafayette, Indiana 47907 USA}

\author{Jukka I. V\"ayrynen}

\affiliation{Department of Physics and Astronomy, Purdue University, West Lafayette, Indiana 47907 USA}

\date{\today}

\begin{abstract}
Fractional changes in entropy are often associated with strong electron correlations or non-Abelian anyons in topological systems. We show that neither condition is necessary -- a system of free electrons on a ring scattering off a repulsive point-like impurity undergoes a $k_B\ln\sqrt 2$ entropy change as the impurity strength is increased. This effect arises from a  shift of allowed single-particle momenta from integer quantization to half-integer quantization, induced by the impurity.
We show that this fractional entropy change is universal across a broad class of single-particle dispersions, demonstrating that such signatures can arise without strong correlations or topological order. 
\end{abstract}

\maketitle

\section{\label{sec:introduction}Introduction}
The emergence of exotic phenomena due to electron correlations is one of the most striking features of many-body physics. Interacting electrons organize into quantum states that support quasiparticle excitations with properties that cannot be inferred from single-particle physics, such as fractional charge \cite{PhysRevLett.50.1395, PhysRevLett.72.724, y9qj-g447} and anyonic statistics \cite{PhysRevLett.53.722, PhysRevB.105.075433, PhysRevLett.125.196802}. Examples include the formation of Cooper pairs in superconducting systems \cite{PhysRev.108.1175}, and fractional quantum Hall effect in two-dimensional electron systems ~\cite{RevModPhys.71.S298, PhysRevLett.50.1395}.
\par
A paradigmatic yet tractable setting to understand how emergent principles are linked to microscopic physics is provided by ``quantum impurity models" \cite{affleck2009quantumimpurityproblemscondensed}, where localized quantum impurities interact with a bath of otherwise free electrons. The simplest such model is the Kondo model, wherein a single magnetic impurity is coupled to an electron bath via an exchange interaction \cite{kondo1964resistance, hewson1997kondo}. Generalizing to multiple channels profoundly alters the physics~\cite{
nozieres1980kondo,affleck1995conformal,1998AdPhy..47..599Z}: in the low-temperature limit, the channels compete to screen the impurity, leading to frustration and the emergence of a non-Fermi liquid fixed point characterized by non-Abelian anyons \cite{PhysRevB.101.085141, affleck1995conformal}. For example, the minimal extension to two channels yields an emergent Majorana fermion, while extension to three channels yields a Fibonacci anyon \cite{doi:10.7566/JPSJ.90.024708}.

A defining feature of non-Abelian anyons is their quantum dimension $d>1$, which reflects the non-trivial degeneracy of the Hilbert space under fusion. In quantum impurity models, the associated entropy is
\begin{equation}
    S= \ln d,
\end{equation}
where we have set $k_B=1$. 
In the two-channel Kondo (2CK) model, the residual impurity entropy in the large coupling limit is independently found to be \cite{affleck1995conformal, doi:10.7566/JPSJ.90.024708, 1998AdPhy..47..599Z, PhysRevB.101.235131, PhysRevB.105.035151, Lotem22, PhysRevB.46.10812, 1994PhRvB..4910020S, Coleman_1995, Rozhkov_1998}
\begin{equation}
    S =\frac{1}{2}\ln 2= \ln \sqrt2,
\end{equation}
which coincides with $\ln d$ for an emergent Majorana mode with $d=\sqrt2$. This entropy which is a logarithm of a non-integer number is often called ``fractional entropy''~\cite{Coleman_1995,Rozhkov_1998,PhysRevLett.123.147702, FENDLEY20091547,
PhysRevLett.128.146803}. 
Recent advances in mesoscopic device technology have made the 2CK effect \cite{2015Natur.526..233I, 2007Natur.446..167P, 2015Natur.526..237K, 2015Natur.526..233I} and fractional entropy directly measurable \cite{PhysRevLett.128.146803, hartman2018direct, PhysRevLett.123.147702, child2022robust, child2021entropy, pyurbeeva2022electronic, piquard2026experimentalevidencefractionalentropy, drayne2026entropicsignaturessingleimpuritykondo}. 
In practice, this fractional entropy is usually measured as an entropy change in the system as Kondo interaction is tuned from off to on. 

These observations have reinforced the prevailing intuition that fractional entropy requires at least one of the two ingredients: strong electron-electron correlations and non-Abelian anyons. In this paper, we show both to be unnecessary, presenting a model of free electrons that exhibits fractional impurity entropy. The paper is organized as follows: in Sec.~\ref{sec:setup}, we present the model of free electrons and obtain its single-particle spectrum. In Sec.~\ref{sec:thermodynamics}, we calculate the partition function, and hence, the impurity entropy of the model. In Sec.~\ref{sec:shifting} we offer an explanation of our result. In Sec. ~\ref{sec:relative-entropy} we demonstrate that this result applies to all dispersions of the form $\epsilon_n=\varepsilon_0n^p$, where $n$ is an integer and $p\geq 1$. Finally, we conclude our discussion in Sec.~\ref{sec:conclusion}.

\begin{figure}[t]
\centering
    \includegraphics[width=\linewidth]{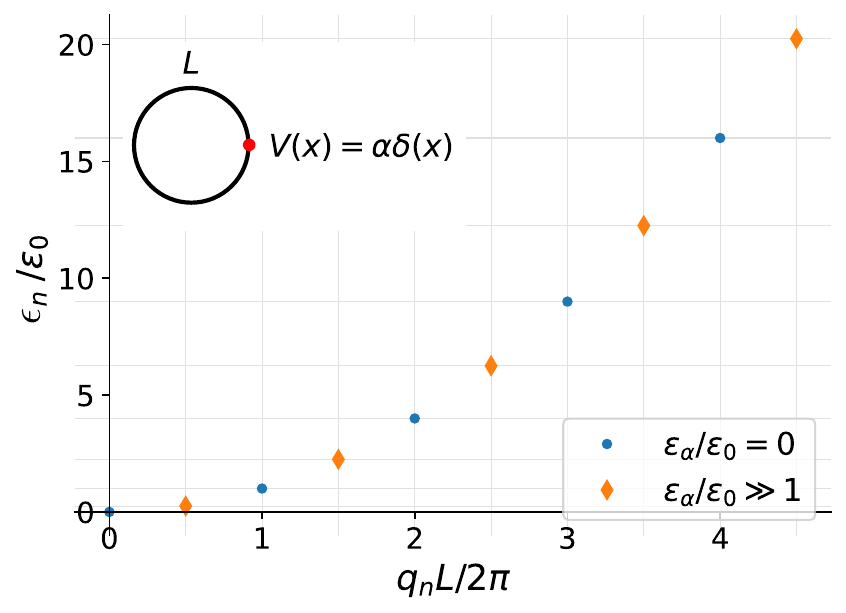}

\caption{Single-particle spectrum of free electrons confined to a one-dimensional ring of circumference $L$ and scattering off a localized, repulsive delta-function potential of strength $\alpha$ (inset). Here, $\varepsilon_\alpha=\alpha/L$ characterizes the impurity strength, while $\varepsilon_0=\pi^2/2mL^2$ is the characteristic finite-size energy scale for the quadratic dispersion given in Eq.~(\ref{eq:unperturbed-dispersion}). As $\varepsilon_\alpha/\varepsilon_0$ increases, the allowed momenta evolve from $q_nL/2\pi=n$ (blue points) to $q_nL/2\pi=n+\tfrac12$ (orange diamonds), where $n=0,1,2,\ldots$. Consequently,   the corresponding single-particle energy spectrum is shifted as well. 
The shifting of the levels leads to a universal entropy change.  
}
\label{fig:model}
\end{figure}

\section{\label{sec:setup} Model}

Consider a system of free electrons confined to a one-dimensional ring of circumference $L$ containing a localized repulsive delta-function scatterer of strength $\alpha$, as illustrated in Fig.~\ref{fig:model}. Then, the single-particle Schrodinger equation is
\begin{equation}
    \left[-\frac{1}{2m}\frac{d^2\psi(x)}{dx^2}+\alpha\delta(x)\right]=E\psi(x).
    \label{eq:hamiltonian}
\end{equation} 
(Throughout, we use units $\hbar = k_B = 1$.) 
In the absence of the impurity,  the (unnormalized) eigenstates are plane waves 
\begin{equation}
    \psi_{n,\pm}(x)= e^{i k_nx} \pm e^{-i k_nx} ,
\end{equation} where  the momenta are quantized as 
\begin{equation} \label{eq:qevenquantization}
k_n=\frac{2n\pi}{L} ,
\end{equation}
  with $n\in\mathbb Z$, following periodic boundary conditions. 
The index $\pm$ denotes parity under $x\to -x$. 
The  single-particle energies are parity-degenerate, 
\begin{equation}
    \epsilon^0_n=\frac{k^2_n}{2m}.\label{eq:unperturbed-dispersion}
\end{equation}

Next,  consider the case of finite $\alpha$, which lifts the spatial parity degeneracy. 
The symmetric eigenstate of  Eq.~(\ref{eq:hamiltonian}),  \begin{equation}
    \psi_{n,+}(x)=\cos(q_n|x|)-\left(\frac{m\alpha}{q_n}\right)\sin(q_n|x|),
\end{equation} 
couples to the scattering potential and has momentum determined by the equation 
\begin{equation}
    q_n = \frac{2n\pi}{L} + \frac{2}{L}\arctan\left(\frac{m\alpha}{ q_n}\right), 
    \label{eq:scattering-momentum-quantisation}
\end{equation} where  $n=0,1,\dots$. 
On the other hand, the anti-symmetric solution
\begin{equation}
    \psi_{n,-}(x)=\sin(p_n x)
\end{equation} 
has a node at the impurity position $x=0$, which means that it does not couple to the impurity, and hence remains identical to the unperturbed case with $p_n=2n\pi/L$ where  $n=1,2,\dots$.

The corresponding single-particle energies are
\begin{equation}
    \epsilon^{\text{S}}_n = \frac{q_n^2}{2m},\label{eq:symmetric-dispersion}
\end{equation}
and
\begin{equation}
    \epsilon^{\text{AS}}_n = \frac{p_n^2}{2m},
    \label{eq:anti-symmetric-dispersion}
\end{equation}
for the symmetric and anti-symmetric branches, respectively. Since the impurity imparts a phase shift to the momentum $q_n$ relative to the unperturbed momentum quantization $k_n$, the symmetric branch is shifted upward with respect to the unperturbed spectrum, as illustrated in Fig.~\ref{fig:model}. The second quantized form of the Hamiltonian, assuming zero chemical potential $\mu$, is
\begin{align}
    H=&
    \sum_{n=0}^\infty \epsilon^{\text{S}}_n(\alpha)\, \hat b^\dagger_q\hat b_q + \sum_{n=1}^\infty\epsilon^{\text{AS}}_n(0) \hat c^\dagger_k\hat c_k.
\end{align} Here, $\hat b_k$ ($\hat b^\dagger_k$) annihilates (creates) an electron corresponding to the symmetric solution, while $\hat c_k$ ($\hat c^\dagger_k$) annihilates (creates) an electron corresponding to the anti-symmetric solution. We have also made the dependence of single-particle spectra on the scattering strength explicit by writing $\epsilon^\mathrm S_q(\alpha)$ and $\epsilon^\mathrm {AS}_p(0)$, emphasizing that only the former is modified by $\alpha$.
\section{Thermodynamics \label{sec:thermodynamics}}
We now investigate the thermodynamic properties of this system via its partition function. This system features three energy scales, namely (i) $\varepsilon_0=\pi^2/2mL^2$, the level spacing at the Fermi level when $\mu=0$; (ii) $\varepsilon_\alpha=\alpha/L$, the energy associated with the scattering potential; and (iii) $T$, the temperature. This gives rise to two independent dimensionless ratios, which we take to be $\varepsilon_0/T$ and $\varepsilon_\alpha/T$. In the following, we analyze the behavior of the impurity entropy $S_\mathrm{imp}$ [defined in Eq.~(\ref{eq:Simp-def})] as a function of $\varepsilon_\alpha/T$ in the limit $\varepsilon_0/T\ll 1$. This limit ensures that the temperature exceeds the level spacing at $\mu=0$, so that several electronic states are thermally accessible for impurity scattering. Furthermore, the only constraint assumed for the temperature is that it is smaller than the bandwidth of the system, so that the dispersion can be approximated as quadratic. 
We will show that the entropy of the system undergoes a decrease of $\tfrac{1}2{\ln2}$ as $\varepsilon_\alpha/T$ is increased from $0$ to  $\infty$. 
\subsection{Impurity partition function}
The partition function of the system at temperature $T=\beta^{-1}$ is 
\begin{align}
    \mathcal Z(\alpha) =&\mathrm {Tr}\, e^{-\beta H}\\ =& \mathrm{Tr}\, \exp{\left[-\beta \left(\sum_{n=0}^\infty \epsilon^{\text{S}}_n(\alpha) \hat b^\dagger_q\hat b_q+\sum_{n=1}^\infty \epsilon^{\text{AS}}_n(0) \hat c^\dagger_k\hat c_k \right)\right]},\label{eq:full-Z}
\end{align}
Since the symmetric and anti-symmetric single-particle states are independent of each other, the partition function factorizes over the two sectors as $\mathcal{Z}(\alpha)=\mathcal Z_\mathrm S(\alpha)\mathcal Z_\mathrm{AS}(0)$,  where 
\begin{align}
    \mathcal{Z}_\mathrm{S}(\alpha)=&\prod_{n=0}^\infty \left(1+e^{-\beta  \epsilon^{\text{S}}_n(\alpha)}\right) \\
\mathcal{Z}_\mathrm{AS}(0)=&\prod_{n=1}^\infty \left(1+e^{-\beta \epsilon^{\text{AS}}_n(0)}\right).
\end{align}

We are interested in determining how the impurity modifies the thermodynamics of the system relative to the unperturbed case. To this end, we introduce an impurity partition function $\mathcal Z_{\text{imp}}$, which is defined as the ratio of the partition function of the full system Eq.~(\ref{eq:full-Z}) with respect to the partition function of the unperturbed system $\mathcal{Z}(0)=\prod_{n=-\infty}^\infty \left(1+e^{-\beta\epsilon_n^0}\right)$:
\begin{align}
\mathcal Z_{\text{imp}}(\alpha)=&\frac{\mathcal Z(\alpha)}{\mathcal Z(0)}.
\end{align}
Note that $\epsilon_n^0 = \epsilon_n^{\text{AS}} = \epsilon_n^{\text{S}}(0)$. 

The antisymmetric sector drops out of the impurity partition function entirely, as expect for the states that do not couple to the impurity. 
We find  
\begin{align}
     \mathcal{Z}_\text{imp}(\alpha)=\prod_{n=0}^\infty\frac{ 1+e^{-\beta  \epsilon^{\text{S}}_n(\alpha)}}{ 1+e^{-\beta \epsilon_n^0}}.
     \label{eq:final-Z}
\end{align}
Consequently, the impurity partition function is determined solely by the  symmetric spectrum which is sensitive to the impurity strength $\alpha$. 
\subsection{Impurity entropy} 
Recall that the thermodynamic entropy $S$ is obtained from the partition function $\mathcal Z$ by
\begin{equation}
    S(\alpha)=\frac\partial{\partial T}[T\ln\mathcal Z(\alpha)]. 
\end{equation} Therefore, the entropy associated with the impurity partition function obtained in Eq.~(\ref{eq:final-Z}) is 
\begin{equation}
    S_\mathrm{imp}(\alpha)=S(\alpha) - S(0)=\frac\partial{\partial T}[T\ln\mathcal Z_\mathrm{imp}(\alpha)], \label{eq:Simp-def}
\end{equation}
i.e., it is also equal to the entropy change of the system upon the introduction of the impurity potential $\alpha$. 

\begin{figure}
    \centering
    \includegraphics[width=0.5\textwidth]{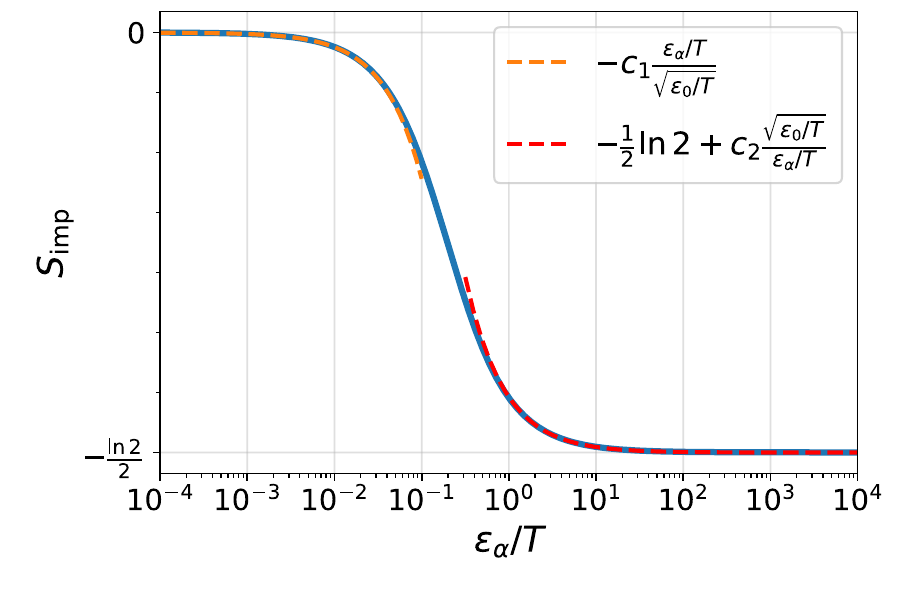}
    \caption{Numerically evaluated impurity entropy (blue curve)  as a function of the scattering potential strength $\varepsilon_\alpha=\alpha/L$ for fixed $\varepsilon_0/T=4.93\times 10^{-2}$. As the potential strength increases, the system undergoes an entropy change of $S_\mathrm{imp}=-\tfrac12\ln2$. The asymptotics obtained in Eq.~(\ref{eq:S-asymptotes}) are also plotted, which show excellent agreement with $S_\mathrm{imp}$ in the small- and large- $\varepsilon_\alpha/T$ limits. }
    \label{fig:entropy}
\end{figure}

We next investigate the asymptotic limits of weak and strong impurity scattering. Assuming that the temperature is much larger than the low-energy level spacing, $T\gg \varepsilon_0$, we find 
\begin{equation}
S_\mathrm{imp}(\alpha)\approx
    \begin{cases}
        -c_1\cdot\frac{\varepsilon_\alpha/T}{\sqrt{\varepsilon_0/T}},\quad &\varepsilon_\alpha/T\ll \varepsilon_0/T \ll 1 \\
        -\frac{\ln2}{2} +c_2\cdot\frac{\sqrt{\varepsilon_0/T}}{\varepsilon_\alpha/T},\quad &\varepsilon_0/T \ll 1 \ll \varepsilon_\alpha/T
    \end{cases},\label{eq:S-asymptotes}
\end{equation}
where $c_1$ and $c_2$ are constants given as
\begin{align}
    c_1&=\frac{\sqrt{\pi}}{4}\left(1-\sqrt 2\right)\zeta\left(\frac{1}{2}\right)&&\approx 0.2681, \\
    c_2&=\frac{3}{2\pi^{3/2}}\left(1-\frac{1}{\sqrt{2}}\right)\zeta\left(\frac{3}{2}\right) &&\approx 0.2061.
\end{align} 
The details of this calculation are presented in Appendix~\ref{sec:entropy-calc} and numerically verified in Fig.~\ref{fig:entropy}. Hence, we conclude that 
$S_\text{imp}\to -\ln 2/2$ when $\varepsilon_\alpha/T\to \infty$. 
\section{Physical interpretation \label{sec:shifting}}
\begin{figure}
    \centering
    \includegraphics[width=1\linewidth]{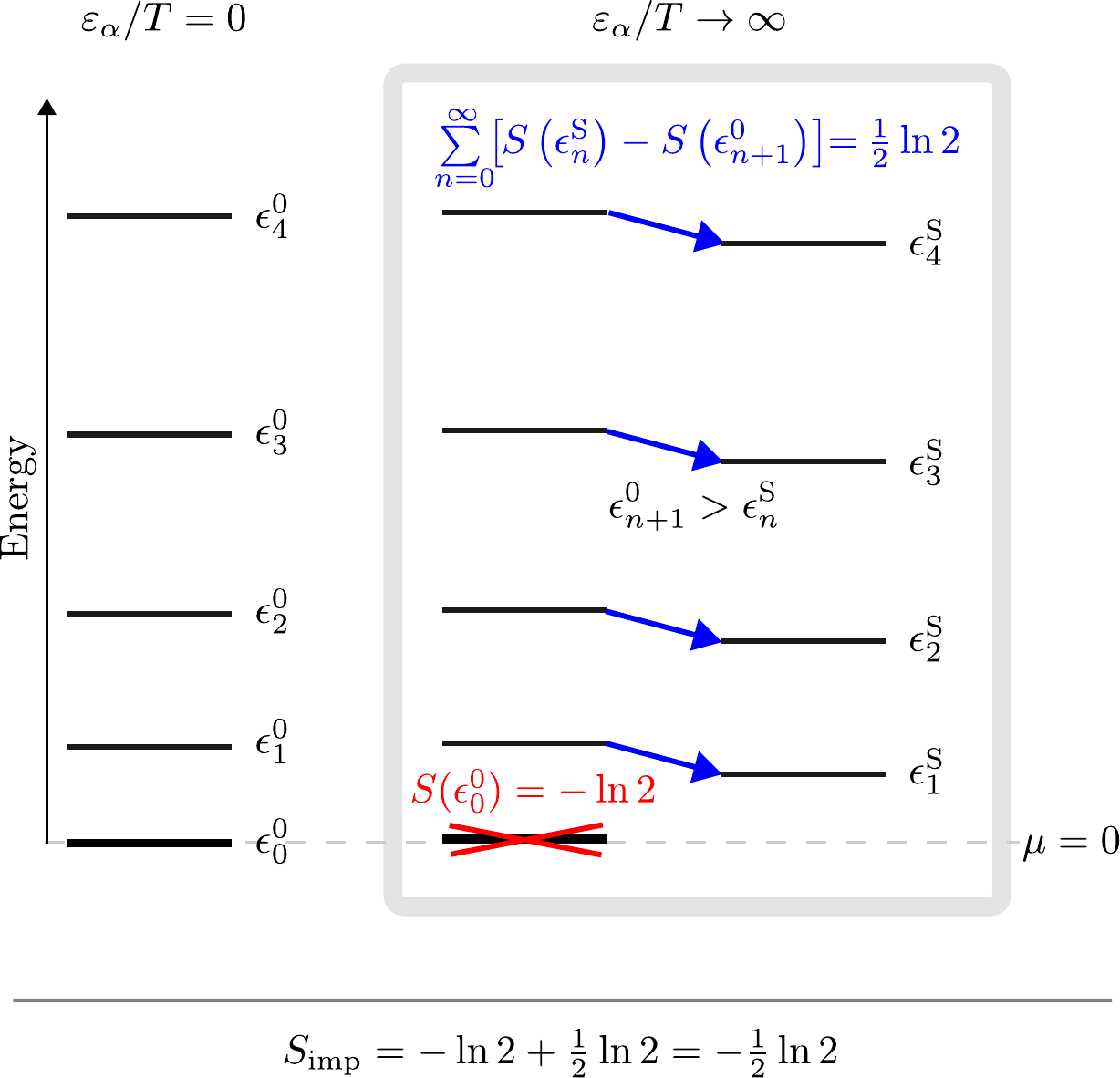}
\caption{Schematic evolution of the energy levels and the associated change in entropy as the scattering potential is changed. \textit{Left}: the energy spectrum with the scattering potential turned off, i.e., $\varepsilon_\alpha/T=0$, with entropy $S(0)=\ln2+\sum_{n=1}^\infty S(\epsilon_n^0)$, the first term coming from ground state degeneracy due to a zero-energy level (note chemical potential $\mu = 0$ here). \textit{Boxed}: Turning on $\varepsilon_\alpha/T$ removes the zero-energy state and shifts each excited  level downwards. 
The former effect lowers the entropy by $\ln 2$ while 
Eq.~(\ref{eq:entropy/0}) shows the latter effect contributes $\sum_{n=0}^\infty\left[S(\epsilon_n^\mathrm{S})-S(\epsilon_{n+1}^0)\right]=\tfrac12\ln2$, so the two factors combine to give $S_\mathrm{imp}=-\ln2+\tfrac12\ln2=-\tfrac12\ln2$.}
\label{fig:spectral-reshuffling}
\end{figure}

\begin{figure}
    \centering
\includegraphics[width=0.5\textwidth]{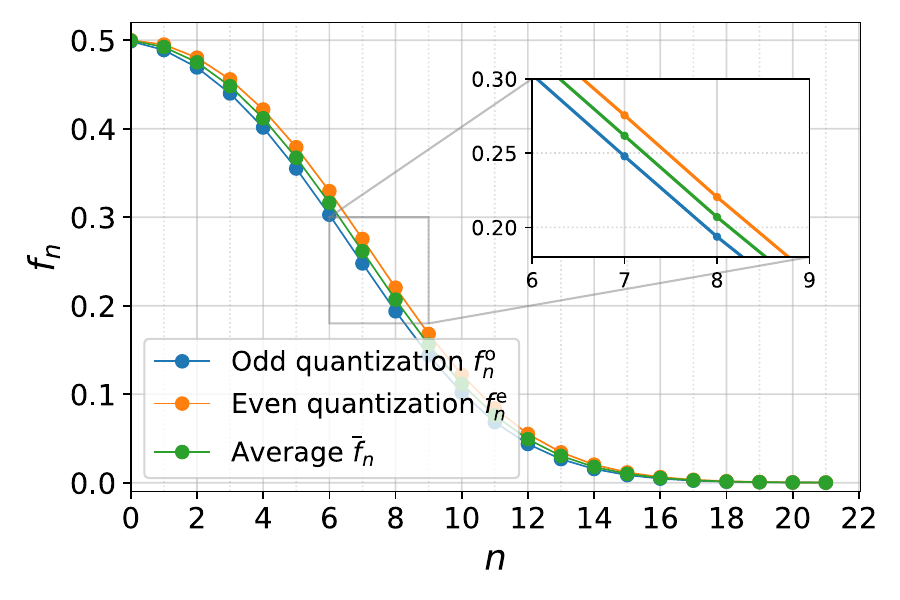}
    \caption{The Fermi distributions $f^\mathrm o_n$ and $f^\mathrm e_n$ of a quadratic spectrum corresponding odd and even momentum quantization condition, respectively. The \textit{average} of these distributions $\overline f_n=(f^\mathrm o_n+f^\mathrm e_n)/2$ is also shown. Here, we fixed $\varepsilon_0/T=4.93\times10^{-3}$.}
    \label{fig:fermi_dist}
\end{figure}
Above, in Fig.~\ref{fig:entropy}, we saw that the impurity entropy changes by a fractional amount $-\frac{1}{2}\ln 2$ as the scattering strength $\alpha$ is changed.  
We will now attempt to develop an intuitive understanding of this result. 
We will see that the entropy change arises from two contributions. 

First, the $\epsilon_0^0=0$ state (corresponding to a uniform wave function) of the unperturbed system results in a  two-fold ground state degeneracy. This degeneracy is removed by $\alpha$ (a uniform state is no longer a solution) and results in a decrease of the entropy by $\ln 2$. 
Second, there is a downward spectral shift of excited states that gives a compensating $+\frac{1}{2}\ln 2$
contribution. 
These two effects are shown schematically in Fig.~\ref{fig:spectral-reshuffling}. Although $\epsilon_n^S > \epsilon_n^0$
for all $n$, a comparison with the truncated unperturbed spectrum 
reveals that $\epsilon_n^S < \epsilon_{n+1}^0$. That is, once the $\epsilon^0_0=0$
state is removed, the scattering levels sit \emph{below} their unperturbed counterparts. Lower-lying states are more thermally accessible and therefore contribute positively to the entropy.

To be concrete, consider the partition function of the truncated (i.e., the zero-energy level removed) unperturbed system,
\begin{equation}
\mathcal{Z}^{/0}_0=\frac{1}{2}\mathcal{Z}_\mathrm S(0),
\end{equation}
where the factor of $1/2$ removes the $n=0$ contribution. The partition function of the scattering system relative to this truncated spectrum is
\begin{equation}
\tilde{\mathcal{Z}}(\alpha)=2\prod_{n=0}^\infty\frac{1+e^{-\beta\epsilon_n^S(\alpha)}}{1+e^{-\beta\epsilon_n^0}}=2\mathcal{Z}_\mathrm{imp}(\alpha),
\end{equation}
where $\mathcal{Z}_\mathrm{imp}(\alpha)$ is defined in Eq.~(\ref{eq:final-Z}). Using Eq.~(\ref{eq:S-asymptotes}) in the large-$\alpha$ limit,
\begin{equation}
\lim_{\varepsilon_\alpha/T\to\infty}\tilde{S}(\alpha)=\ln 2-\frac{\ln 2}{2}=+\frac{\ln 2}{2}.\label{eq:entropy/0}
\end{equation}
This is precisely the missing entropy. The two contributions to entropy, namely $-\ln 2$ from the removal of the zero-energy level and $+\frac{1}{2}\ln 2$ from spectral shifting, sum to $S_\mathrm{imp}(\alpha)=-\frac{1}{2}\ln 2$, confirming that fractional entropy can arise purely from a shifting of single-particle levels, with no exotic statistics required.

In fact, this result is not tied to impurity scattering. Consider Eq.~(\ref{eq:scattering-momentum-quantisation}): as noted previously, scattering shifts the allowed momenta upward with respect to their unperturbed counterparts. This shift is maximal when $m\alpha/q_n\to\infty$, for which
\begin{equation}\label{eq:qoddquantization}
q_n=\frac{\pi}{L}(2n+1),
\end{equation}
so that the momenta are quantized by odd integers rather than even integers as in the unperturbed problem. Consequently, in the large-$\varepsilon_\alpha/T$ limit of Eq.~(\ref{eq:S-asymptotes}), $S_\mathrm{imp}(\alpha)$ measures precisely the entropy difference between free spectra with odd, Eq.~(\ref{eq:qoddquantization}), and even, Eq.~(\ref{eq:qevenquantization}), momentum quantization. Motivated by this observation, we define the ``boundary entropy change''
\begin{equation}
\Delta S\equiv\lim_{\varepsilon_\alpha/T\rightarrow\infty}S_{\mathrm{imp}}(\alpha)=S_{\mathrm{odd}}-S_{\mathrm{even}}.
\label{eq:DeltaS-def}
\end{equation}
Thus, $\Delta S$ is not just tied to the impurity problem, but instead characterizes the entropy difference associated solely with the half-integer offset between the two momentum quantization conditions. In the next section, we evaluate $\Delta S$ for a broad class of dispersions, demonstrating that the fractional entropy change is a universal consequence of this shifting of  momenta. 

\section{Universality of the boundary entropy change}
\label{sec:relative-entropy}
Let us consider a general dispersion of the form $\epsilon_n=\varepsilon_0 n^p$ with $n=0,1,2,\dots$ labeling the quantized momenta and assume $\varepsilon_0/T \to0$. The exponent  $p\geq1$ characterizes the energy-momentum dispersion relation. 
We now show that the fractional entropy change is a universal feature of the system when the momentum quantization changes by a half-integer. 
Specifically, the momenta quantized by odd and even integers correspond to respective  energies 
\begin{equation}
    \epsilon^\mathrm o_n= \varepsilon_0 (2n+1)^p,
 \quad   \epsilon^\mathrm e_n=\varepsilon_0 (2n)^p.
\end{equation}

The entropy can be expressed in the standard form
\begin{equation}
    S=-\sum_{n=0}^\infty \left[f_n\ln f_n+(1-f_n)\ln(1-f_n)\right],
    \label{eq:entropy-fermi-def}
\end{equation}
where $f_n$ is the Fermi weight of the $n$-th mode, evaluated on the respective spectrum. 
Henceforth, for brevity, all distribution functions constructed using the odd (even) momentum quantization will be referred to as odd (even) distributions $f_n^{\mathrm{o}}$ ($f_n^{\mathrm{e}}$).

Fig.~\ref{fig:fermi_dist} shows the odd and even Fermi distributions. 
In the thermodynamic limit $\varepsilon_0 / T \ll 1$, these distributions remain close to each other. To see this, define the \textit{average dispersion} $\bar\epsilon_n=(\epsilon^\text{o}_n+\epsilon^\text{e}_n)/2$, so that the odd and even dispersions take the form
\begin{equation}
    \epsilon^\mathrm{o,e}_n=\bar\epsilon_n\pm \Delta\epsilon_n,
\end{equation}
where $\Delta\epsilon_n=(\epsilon^\mathrm{o}_n-\epsilon^\mathrm{e}_n)/2$ measures the deviation of the odd/ even spectra from the average. The odd and even dispersions are close whenever $\Delta\epsilon_n$ is small for the thermally relevant energy levels, i.e., $\Delta \epsilon_n/\bar\epsilon_n\approx p/n$ is small. At temperature $T$, the thermally relevant energy levels are $\bar\epsilon_n\approx T$, which gives $n\approx (T/\varepsilon_0)^{\frac1p}$. Hence, $\Delta \epsilon_n/\bar\epsilon_n\approx p/n\approx p(\varepsilon_0/T)^\frac1p\ll 1$ whenever $\varepsilon_0/T\ll1$. Under this condition, the odd and even Fermi distributions are therefore close, and we may expand 
\begin{equation}
    f^\mathrm{o,e}(\bar\epsilon_n\pm\Delta\epsilon_n)\approx f(\bar\epsilon_n)\mp f^\prime(\bar\epsilon_n)\Delta\epsilon_n.
\end{equation} 

The boundary entropy change $\Delta S$ can therefore be expanded in powers of $\Delta \epsilon_n$. To the lowest order, we have (see Appendix~\ref{sec:entropy-approx-calc})
\begin{align}
    \Delta S\approx-\frac12\sum_{n=0}^\infty\left[\frac{\bar\epsilon_n}{T^2}\mathrm{sech}^2\left(\frac{\bar\epsilon_n}{2T}\right)\right]\Delta\epsilon_n.\label{eq:entropy-approx-02}
\end{align}
Note that this result is entirely general in that no assumptions about the form of the dispersion were made to derive it. Next, we will evaluate the sum in Eq.~(\ref{eq:entropy-approx-02}) to obtain the boundary entropy change between odd and even dispersions of any dispersion of the form $\epsilon_n=\varepsilon_0 n^p$.

Note that the sum Eq.~(\ref{eq:entropy-approx-02})  is reminiscent of a Riemann sum; however, $\Delta\epsilon_n$ is not the correct measure here, for it captures the deviation of the odd/even spectra from the average dispersion. The proper  measure would be $\Delta\bar\epsilon_n=\bar\epsilon_{n+1}-\bar\epsilon_n$, the difference between two consecutive levels in the average dispersion. Hence, a relationship between $\Delta\epsilon_n$ and $\Delta\bar\epsilon_n$ is sought. We show in Appendix.~\ref{sec:delta-eps-delta-eps-bar} that
\begin{equation}
    \lim_{n\to\infty}\frac{\Delta\epsilon_n}{\Delta \bar\epsilon_n}=\frac14\label{eq:eps-eps-bar}.
\end{equation}
Using this result in Eq.~(\ref{eq:entropy-approx-02}) we obtain a Riemann sum: 
\begin{align}
    \Delta S&\approx -\frac{1}{2}\sum_{n=0}^{\infty}\left[\frac{\bar{\epsilon}_{n}}{4T^{2}}\mathrm{sech}^{2}\left(\frac{\bar{\epsilon}_{n}}{2T}\right)\right]\Delta\bar{\epsilon}_{n}.\label{eq:DeltaS-approx-sum}
\end{align}

In the limit of large $T/\varepsilon_0$, and therefore large $n$, we have 
 $\Delta\bar\epsilon_n / \bar{\epsilon}_n\to 0$ allowing  us to convert the sum into an integral
 \begin{align}
    \Delta S &\approx-\frac{1}{2}\int_{0}^{\infty}\left[\frac{\bar{\epsilon}}{4T^{2}}\mathrm{sech}^{2}\left(\frac{\bar{\epsilon}}{2T}\right)\right]\,d\bar{\epsilon}\\
    &= -\frac{1}{2}\int_{0}^{\infty}x\mathrm{sech}^{2}\left(x\right)\,dx \\
    &=-\frac{\ln2}{2}.
\end{align}

Hence, we have found that $\Delta S\approx-\ln2/2$, which is remarkable in two ways. First, it is entirely general: no assumptions about the form of the dispersion were made, rendering it applicable to all dispersions of the form $\epsilon_n=\varepsilon_0 n^p$, for $\varepsilon_0/T \to0$ and $p\geq1$. Second, the calculation relies on the smoothness of the Fermi distributions $f^\mathrm{o}$ and $f^\mathrm e$, which is ensured when $\varepsilon_0/T<1$. Once this condition is satisfied, the result is independent of temperature, since the integral carries no explicit temperature dependence. Both facts underscore that $\Delta S$ depends only on the relative quantizations of the two spectra, and not their detailed structure.

An illustrative example in this case is of chiral Majorana modes $\hat\gamma_k$ on a ring~\cite{bernevig2015topologicalsuperconductorscategorytheory}: $\hat H=\sum_{k>0} \epsilon_k\hat\gamma_{-k}\hat\gamma_k$, where $\epsilon_k=vk$ is a linear dispersion. Threading the ring with a flux $\Phi$ changes the boundary conditions and the momentum quantization as 
\begin{equation}
    k_n=\frac{2n\pi}{L}+\frac{\theta} {L},
\end{equation}
where $\theta=\pi\Phi/\Phi_0$, and $\Phi_0=hc/2e$ is the superconducting flux quantum. 
The cases $\theta=0$ and $\theta=\pi$ correspond to even and odd momentum quantizations, respectively, and we find $\Delta S=-\tfrac12\ln 2$. This result is analogous to the flux-induced transition between even and odd Majorana sectors in a two-dimensional $p$-wave superconductor. The even sector contains a Majorana zero mode, which is lost as the momentum quantization changes to odd as a consequence of flux threading. This reflects in a change of entropy $\Delta S=-\tfrac12\ln2$.
\subsection{Boundary entropy change at finite chemical potential}

\begin{figure}[h]
    \centering
        \includegraphics[width=\linewidth]{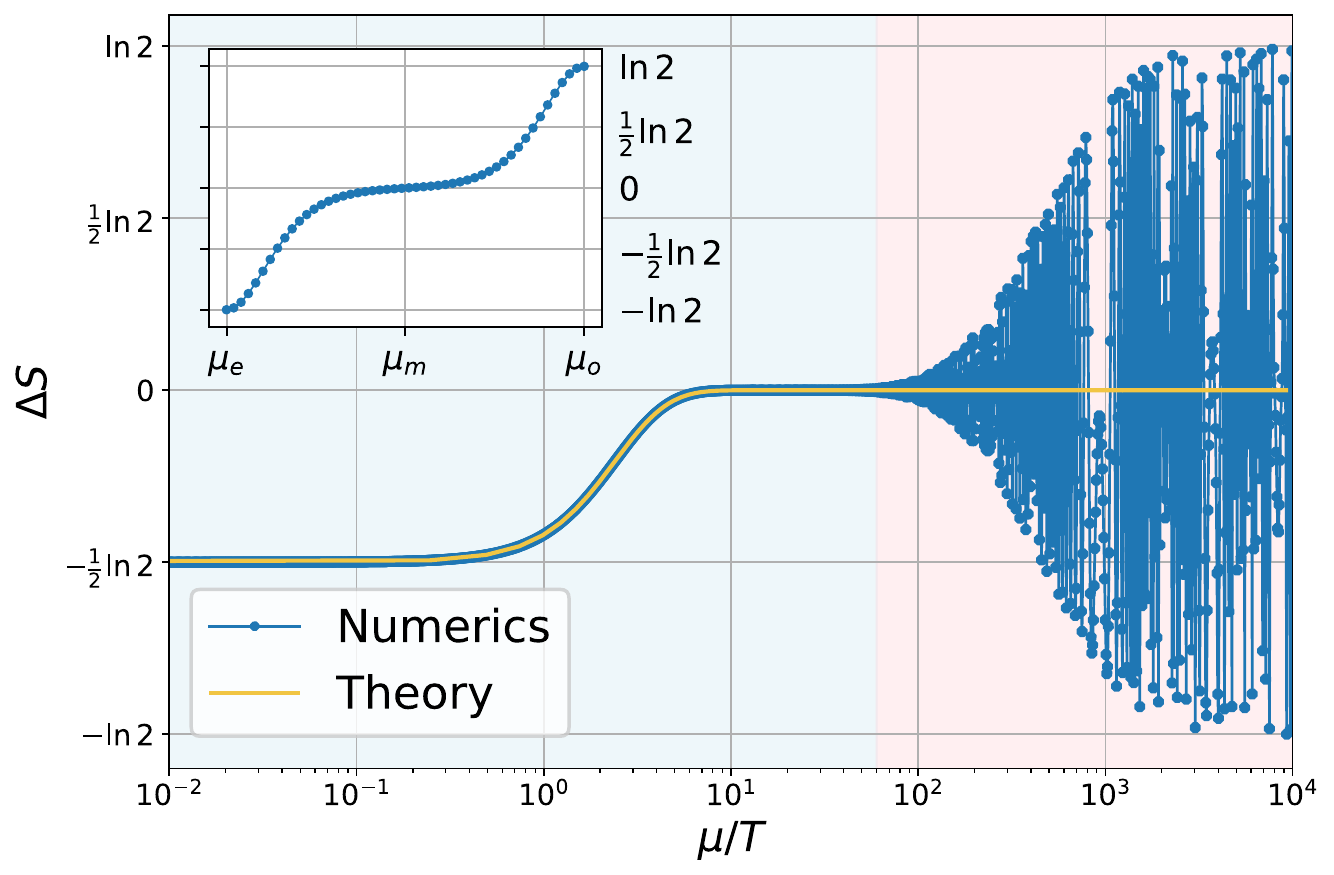}
        \caption{Boundary entropy change of a quadratic dispersion ($p=2$) is plotted as a function of chemical potential $\varepsilon_0 /T=4.93\times10^{-3}$. Two distinct regimes are visible. In the blue region, $T/\varepsilon_\mu > 1$: temperature is larger than the level spacing near the Fermi level, so many energy states are thermally accessible (``continuum limit''). Initially, when $\mu$ is small, not many levels before the Fermi level contribute to the entropy, so $\Delta S\approx -\ln2/2$ as we saw earlier. However, as $\mu$ grows, more levels are occupied and $\Delta S$ grows toward zero. The analytical result (red curve) obtained in Eq.~(\ref{eq:Delta_S_mu_dependent}) shows excellent agreement with the numerical result in this region.  In the pink region, $T/\varepsilon_\mu\ll 1$: since $\Delta\bar\epsilon_n$ grows with $n$ for all $p>1$, the level spacing eventually exceeds the  temperature for large enough $\mu$, and we approach an effective zero-temperature limit (``discrete limit''). In this regime, only energy levels close to the Fermi level contribute to the entropy and so $\Delta S$ begins to rapidly oscillate with $\mu$. These oscillations are bounded by $\Delta S=\pm \ln2$, with the bounds saturated when an odd or even level coincide with the Fermi level. The inset shows magnified view of the discrete regime, highlighting a single $\Delta S$ oscillation as chemical potential increases from $\mu=\epsilon^\text{e}_n$ to $\mu=\epsilon^\text{o}_n$ for $n=225$.}
        \label{fig:entropy_mu_dependence}
\end{figure}

\begin{figure}[h]
    \centering
        \centering
        \includegraphics[width=\linewidth]{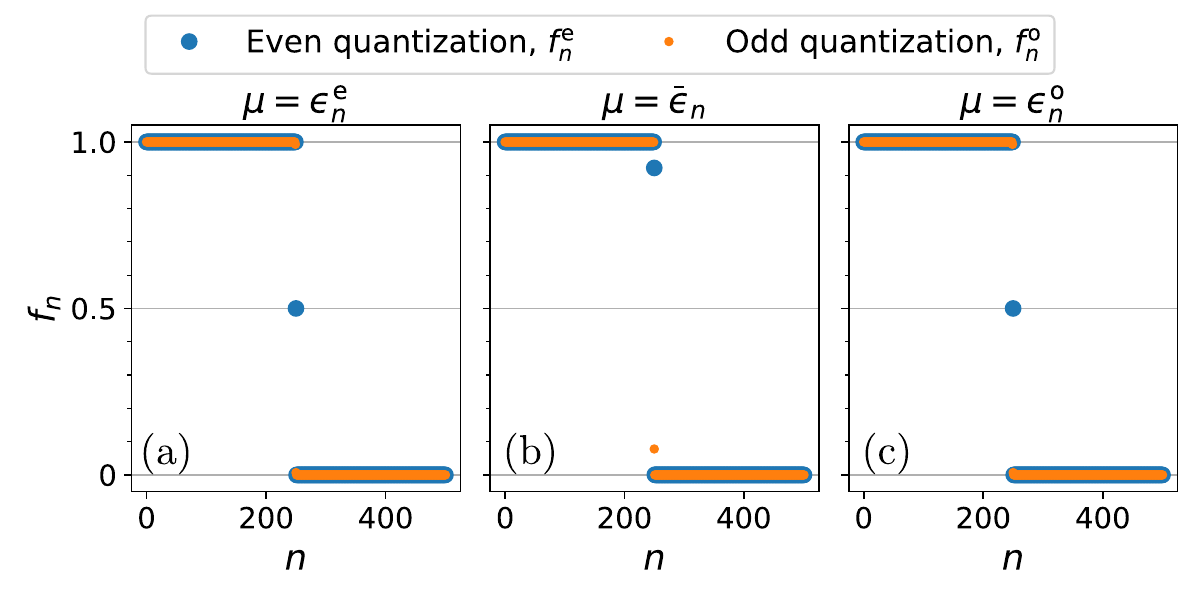}
    \label{fig:fermi_distributions_comparison}
    \caption{The Fermi distributions of  a quadratic dispersion for three values of the chemical potential are plotted for $\varepsilon_0/T=4.93\times10^{-3}$. 
     In panels (a), (b), and (c), the chemical potential is fine-tuned to intersect with the even, average, and odd energy level associated with $n=225$. 
    For this $n$, the level spacing is large enough for the system to be in the effective zero temperature limit, so the Fermi distribution behaves like a step-function.}
    \label{fig:right_zoomed}
\end{figure}
So far, our discussion has focused on the case of zero chemical potential. We now consider a finite chemical potential $\mu>0$, for which the single-particle energies become $\epsilon_n=\varepsilon_0n^p-\mu$, and investigate its effect on $\Delta S$. The resulting behavior is shown in Fig.~\ref{fig:entropy_mu_dependence}, where $\Delta S$ is plotted as a function of $\mu$. As $\mu$ increases, the Fermi level moves deeper into the spectrum, where the level spacing increases as a function of $n$ for $p>1$. Therefore, the behavior of $\Delta S$ is governed by a competition between the thermal fluctuations, which tend to smear out the discreteness of levels, and the increasing level spacing, which resolves the discreteness. Hence, if $\varepsilon_\mu$ denotes the level spacing at the Fermi level at a given chemical potential $\mu$, the result reveals two distinct regimes governed by the ratio $T/\varepsilon_\mu$, which we will analyze below. 
\subsubsection{The continuum limit, $T/\varepsilon_\mu \gg 1$}
As aforementioned, the calculation in Sec.~\ref{sec:relative-entropy} requires the Fermi distribution to be a smooth function, which is a feature of the continuum limit $T/\varepsilon_\mu\gg 1$. However, we find that the approximation remains valid well beyond the asymptotic limit, and only breaks down for sufficiently small $T/\varepsilon_\mu$. This allows us to extend the $\mu=0$  calculation to account for finite chemical potential, which we do by making the replacing $\bar\epsilon_n\to\bar\epsilon_n-\mu$ in Eq.~(\ref{eq:DeltaS-approx-sum})
\begin{align}
    \Delta S&\approx-\frac{1}{2}\sum_{n=0}^{\infty}\left[\frac{\bar{\epsilon}_{n}-\mu}{4T^{2}}\mathrm{sech}^{2}\left(\frac{\bar{\epsilon}_{n}-\mu}{2T}\right)\right]\Delta\bar{\epsilon}_{n} \\
    &\to \frac{1}{2}\int_{-\mu/2T}^\infty x\mathrm{sech}^{2}\left(x\right)\,dx \\
    &\approx \frac{1}{2}\left(\frac{\mu}{2T}\right)\tanh\left(\frac{\mu}{2T}\right) - \frac{1}{2}\ln\left[\cosh\left(\frac{\mu}{2T}\right)\right]. \label{eq:Delta_S_mu_dependent}
\end{align}
Excellent agreement between this approximation and the numerical data is shown in Fig.~\ref{fig:entropy_mu_dependence}.

\subsubsection{The discrete limit, $T/\varepsilon_\mu \ll 1$}
As $\mu/T$ is further increased, we eventually enter the regime where temperature is much less than the level spacing at Fermi level, $T/\varepsilon_\mu \ll 1$,  
and the Fermi distribution becomes step-like. Eventually, only energy levels in the immediate vicinity of the Fermi level contribute appreciably to the entropy, since this is effectively a zero-temperature regime in which levels far from $\mu$ are either fully occupied or fully empty,  thus contributing negligibly. Whether the dominant contribution to $\Delta S$ is positive or negative depends on the level that lies the closest to $\mu$: $\Delta S>0$ when it is a level of the odd spectrum, and vice versa. This produces an oscillation in $\Delta S$ as $\mu/T$ increases and sweeps past successive levels as shown in Fig.~\ref{fig:entropy_mu_dependence}. The amplitude of the oscillations grows as we move to larger $\mu$, for as $\mu$ approaches a level of a select parity, the levels of the other parity necessarily lie farther away, contributing negligible entropy. This is illustrated in Fig.~\ref{fig:right_zoomed}. The amplitude of oscillations are bounded by $\Delta S = \pm \ln2$, with the bounds saturated in the $T/\varepsilon_\mu\to 0$ limit only, when $\mu$ intersects with an odd and even level, respectively, resulting in a two-fold ground state degeneracy. The boundary entropy change also periodically returns to $\Delta S=0$ for $\mu=\bar\epsilon_n$. These values of $\mu$  lie  exactly between some odd and even energy levels (Fig.~\ref{fig:right_zoomed}b), resulting in their contributions canceling out in $\Delta S$. 
\section{Conclusion}
\label{sec:conclusion}
Fractional entropy has long been considered a hallmark of non-Abelian anyons or strong correlations. However, in this work, we have shown that it may arise in a simple non-topological system of free electrons,  
thus demonstrating that fractional impurity entropy is not always be indicative of exotic physics. 
We find that different momentum quantization conditions (dictated by  boundary conditions) lead to a shift of energy levels,  producing a  difference $\Delta S$ in entropies. In particular, we have found that a half-integer shift between momentum quantizations yields $\Delta S=\tfrac12\ln2$ (the entropy associated with Majorana fermions) for dispersions of the form $\epsilon_k \propto k^p$ 
with $p\geq1$. 
This entropy change is how impurity entropy is usually measured in practice, by comparing the entropies with and without Kondo interaction. 
In this work we have shown  that the detection of an entropy change $\Delta S=\tfrac12\ln 2$ should not be regarded as a smoking gun diagnostic for Majorana fermions, and additional confirmation must be sought.

\begin{acknowledgments}
This work was supported by the National Science Foundation award number DMR-2540313. 
\end{acknowledgments}

\appendix
\section{Asymptotic Expansion of the Impurity Entropy}
\label{sec:entropy-calc}
The momentum quantization condition, Eq.~(\ref{eq:scattering-momentum-quantisation}), written in terms of the energy scales $\varepsilon_\alpha$ and $\varepsilon_0$, reads
\begin{equation}
    q_n=\frac{2n\pi}{L}+\frac2L\arctan\left(\frac{\pi^{2}}{2Lq_{n}}\,\frac{\varepsilon_{\alpha}}{\varepsilon_{0}}\right).\label{eq:momentum-quantisation-ee}
\end{equation}
In what follows, we derive the impurity entropy $S_\mathrm{imp}$ as presented in Eq.~(\ref{eq:S-asymptotes}) from the impurity partition function of Eq.~(\ref{eq:final-Z}), using the limiting behavior of the momenta $q_n$.
\subsection{Limit  $\varepsilon_{\alpha}/T\ll\varepsilon_{0}/T\ll1$}
In this limit, the argument of $\arctan$  in Eq.~(\ref{eq:momentum-quantisation-ee}) is small and 
\begin{equation}
    q_n\approx\frac{2n\pi}{L}+\frac{\pi^{2}}{L^{2}q_{n}}\frac{\varepsilon_{\alpha}/T}{\varepsilon_{0}/T}.
\end{equation}
This equation can be readily solved to obtain $q_n$. For $n=0$, we find
\begin{equation}
    q_0\approx\frac{\pi}{L}\sqrt{\frac{\varepsilon_{\alpha}/T}{\varepsilon_{0}/T}},
\end{equation}
while for $n=1,2,\dots$, we obtain
\begin{equation}
    q_n\approx \frac{2n\pi}{L}+\frac{\pi}{2nL}\frac{\varepsilon_\alpha/T}{\varepsilon_0/T}.
\end{equation}
The corresponding energies are obtained using $\epsilon^\mathrm S_n=q_n^2/2m$
\begin{equation}
    \epsilon_n^\mathrm S(\alpha)\approx \begin{cases}
        \varepsilon_\alpha,\quad &n=0, \\
        \epsilon_{n}^{0}+2\varepsilon_{\alpha}, \quad &n=1,2,\dots
    \end{cases}.
\end{equation}
Here $\epsilon_{n}^{0}=\varepsilon_{0}\left(2n\right)^{2}$ is the energy when $\alpha = 0$, see Eq.~(\ref{eq:unperturbed-dispersion}). 
Therefore, the impurity partition function, Eq.~(\ref{eq:final-Z}), is
\begin{align}
    \mathcal{Z}_{\text{imp}}(\alpha)&=\frac{\left(1+e^{-\epsilon_{0}^{\text{S}}/T}\right)\prod_{n=1}^{\infty}\left(1+e^{-\epsilon_{n}^{\text{S}}/T}\right)}{\prod_{n=0}^{\infty}\left(1+e^{-\epsilon_{n}^{0}/T}\right)}\\
    &\approx\frac{\left(1+e^{-\varepsilon_{\alpha}/T}\right)}{2}\prod_{n=1}^{\infty}\frac{1+e^{-\epsilon_{n}^{0}/T}e^{-2\varepsilon_{\alpha}/T}}{1+e^{-\epsilon_{n}^{0}/T}}.
\end{align} 
Next, we expand in small $\varepsilon_\alpha/T$, so that

\begin{equation}
   \frac{1+e^{-\epsilon_{n}^{0}/T}e^{-2\varepsilon_{\alpha}/T}}{1+e^{-\epsilon_{n}^{0}/T}} \approx 1-2f\left(\epsilon_{n}^{0}\right)\frac{\varepsilon_{\alpha}}T,
\end{equation}
where $f\left(\epsilon_{n}^{0}\right)=\left[1+\exp(\epsilon_{n}^{0}/T)\right]^{-1}$. Then, 
\begin{align}
    \ln \mathcal Z_\mathrm{imp}\approx& -\ln2+\ln\left(1+e^{-\varepsilon_{\alpha}/T}\right)+\sum_{n=1}^{\infty}\ln\left[1-2f\left(\epsilon_{n}^{0}\right)\frac{\varepsilon_{\alpha}}T\right]\\
    \approx &-\ln 2 + \ln\left[1+\left(1-\frac{\varepsilon_\alpha} T\right)\right]+\sum_{n=1}^{\infty}\ln\left[1-2f\left(\epsilon_{n}^{0}\right)\frac{\varepsilon_{\alpha}}T\right]\\
    \approx& -\left[\frac{1}{2}+\sum_{n=1}^{\infty}2f\left(\epsilon_{n}^{0}\right)\right]\frac{\varepsilon_{\alpha}}T.\label{eq:small-Zimp-01}.
\end{align} 
Since $\varepsilon_0/T$ is small, $f\left(\epsilon_{n}^{0}\right)$ is a smooth function, and so we can evaluate the sum in Eq.~(\ref{eq:small-Zimp-01}) using the Euler-MacLaurin summation formula \cite{AbramowitzStegun1964}
\begin{equation}
   \sum_{n=1}^\infty \frac{1}{1+e^{(2n)^2\varepsilon_0/T}} \approx \int_0^\infty \frac{dx}{1+e^{(4\varepsilon_0/T)x^2}},\end{equation}
where the corrections to this result are of the order of $\varepsilon_0/T$.

The integral is of standard Fermi-Dirac type and it evaluates to \cite{Dingle1957}
\begin{equation}
    \int_0^\infty \frac{dx}{1+e^{(4\varepsilon_0/T)x^2}}=\left[\frac{\sqrt{\pi}}{4}\left(1-\sqrt{2}\right)\zeta\left(\frac{1}{2}\right)\right]\frac{1}{\sqrt{\left(\varepsilon_{0}/T\right)}}.
\end{equation}
$\zeta$ denotes the  Riemann zeta function. 
Hence, 
\begin{equation}
    \ln\mathcal Z_\mathrm{imp}\approx-\left[\frac{\sqrt{\pi}}{2}\left(1-\sqrt{2}\right)\zeta\left(\frac{1}{2}\right)\frac{1}{\sqrt{\varepsilon_{0}/T}}\right]\frac{\varepsilon_{\alpha}}T+\dots,
\end{equation}

The entropy $S_\mathrm{imp}$ is straightforwardly calculated using thermodynamic relations. We find
\begin{equation}
    S_\mathrm{imp}\approx -\frac{\sqrt{\pi}}{4}\left(1-\sqrt{2}\right)\zeta\left(\frac{1}{2}\right)\frac{\varepsilon_{\alpha}/T}{\sqrt{\varepsilon_{0}/T}} + \dots,
\end{equation}
which is the first limit of  Eq.~(\ref{eq:S-asymptotes})  in the main text. 
\subsection{Limit  $\varepsilon_0/T\ll1\ll\varepsilon_\alpha/T$}
In this limit,  the argument of $\arctan$  in Eq.~(\ref{eq:momentum-quantisation-ee}) is large and  
\begin{equation}
    q_n\approx \frac{\pi}{L}\left(2n+1\right)-\frac{4q_{n}}{\pi^{2}}\frac{\varepsilon_{0}/T}{\varepsilon_{\alpha}/T},
\end{equation}
which gives
\begin{equation}
    q_n\approx \frac{\pi}{L}\left(2n+1\right)\left(1-\frac{4}{\pi^{2}}\frac{\varepsilon_{0}/T}{\varepsilon_{\alpha}/T}\right),
\end{equation}
for $n=0,1,2\dots$ The single-particle energies are
\begin{equation}
    \epsilon_n^\mathrm{S}\approx \varepsilon_{0}\left(2n+1\right)^{2}\left(1-\frac{8}{\pi^{2}}\frac{\varepsilon_{0}/T}{\varepsilon_{\alpha}/T}\right).
\end{equation}
We can, therefore, write the impurity partition function
\begin{align}
    \mathcal Z_\mathrm{imp}(\alpha)\approx\prod_{n=0}^\infty\frac{1+\exp{\left[-\frac{\epsilon_0}{T}(2n+1)^2\left(1-\frac8{\pi^2}\frac{\varepsilon_0/T}{\varepsilon_\alpha/T}\right)\right]}}{1+e^{-\epsilon_n^0/T}}.
\end{align}
Let us denote $c=\varepsilon_0/T$ and $\delta= \tfrac8{\pi^2}\tfrac{(\varepsilon_0/T)^2}{\varepsilon_\alpha/T}$, where $\delta\ll c\ll 1$. Then,
\begin{align}
    \ln\mathcal Z_\mathrm{imp}(\alpha)&\approx \sum_{n=0}^\infty\ln\left[1+e^{(\delta-c)(2n+1)^2}\right]\nonumber\\&-\sum_{n=0}^\infty\ln\left[1+e^{-c(2n)^2}\right].\label{eq:large-Zimp-01}
\end{align} The sums in Eqs.~(\ref{eq:large-Zimp-01}) are written as integrals using the Euler-MacLaurin summation formula. Those integrals are again of standard Fermi-Dirac type and known. We find up to corrections of the order of $\varepsilon_0/T$
\begin{align}
    \sum_{n=0}^\infty \ln\left[1+e^{(\delta-c)(2n+1)^2}\right]&\approx \frac{c_0}{\sqrt{c-\delta}}\\&\approx \frac{c_0}{\sqrt c}\left(1+\frac12\frac{\delta} {c}\right),\label{eq: app-a-inter-01}\\
    \sum_{n=0}^\infty \ln\left[1+e^{-c(2n)^2}\right]&\approx \frac{c_0}{\sqrt{c}}+\frac12\ln2,
\end{align}
where
\begin{align}
    c_0=&\frac{\sqrt\pi}{4}\left(1-\frac{1}{\sqrt{2}}\right)\zeta\left(\frac{3}{2}\right).
\end{align}
Hence, $\ln\mathcal Z_\mathrm{imp}\approx -\tfrac12\ln2+(4c_0/\pi^2)\tfrac{\sqrt{\varepsilon_0/T}}{\varepsilon_\alpha/T}+\dots$, and the impurity entropy is
\begin{equation}
    S_\mathrm{imp}\approx -\frac12\ln2+c_2\frac{\sqrt{\varepsilon_{0}/T}}{\varepsilon_{\alpha}/T}+\dots,
\end{equation}
where $c_2=6c_0/\pi^2$. 
This is the second limit of  Eq.~(\ref{eq:S-asymptotes})  in the main text. 

\section{Approximating boundary entropy change}
\label{sec:entropy-approx-calc}
In this appendix, we will derive the entropy change $\Delta S$, Eq.~(\ref{eq:entropy-approx-02}) of the main text. Let $\bar f_n\equiv f(\bar \epsilon_n)$ denote the average Fermi distribution and $\Delta f_n\equiv f^\prime(\bar\epsilon_n)\Delta\epsilon_n$ the deviation of the odd/even Fermi distributions from the average $\bar f_n$. Then, 
the odd and even entropies $S^\mathrm{o,e}=\sum_{n=0}^\infty S^\mathrm{o,e}_n$ can be expanded about $\bar f_n$ as
\begin{equation}
    S_{n}^{\text{o,e}}\left(\bar f_n\mp\Delta f_n\right)\approx S(\bar f_n)\pm S^\prime(\bar f_n)\Delta f_n.
\end{equation}
The boundary entropy change $\Delta S=S^\mathrm{o}-S^\mathrm{e}$ therefore is
\begin{equation}
    \Delta S\approx2\sum_{n=0}^{\infty}S^\prime(\bar f_n)f^{\prime}\left(\bar{\epsilon}_{n}\right)\Delta\epsilon_{n}.
\end{equation}
Using Eq.~(\ref{eq:entropy-fermi-def}) we find $S^\prime(\bar f_n)=\ln(1/\bar f_n-1)=\bar\epsilon_n/T$, and 
\begin{align}
    f^\prime(\bar\epsilon_n)&=-\frac{1}{T}\frac{e^{\bar\epsilon_{n}/T}}{\left(1+e^{\bar\epsilon_{n}/T}\right)^{2}}\\
    &=-\frac{1}{4T}\mathrm{sech}^{2}\left(\frac{\bar{\epsilon}_{n}}{2T}\right).
\end{align}
Hence, to the first order in $\Delta \epsilon_n$, the boundary entropy change is
\begin{align}
    \Delta S&\approx2\sum_{n=0}^{\infty}\left(\frac{\bar{\epsilon}_{n}}{T}\right)\left[-\frac{1}{4T}\mathrm{sech}^{2}\left(\frac{\bar{\epsilon}_{n}}{2T}\right)\Delta\epsilon_{n}\right]\\
    &=-\frac{1}{2}\sum_{n=0}^{\infty}\left[\frac{\bar{\epsilon}_{n}}{T^{2}}\mathrm{sech}^{2}\left(\frac{\bar{\epsilon}_{n}}{2T}\right)\right]\Delta\epsilon_{n},
\end{align}
which is Eq.~(\ref{eq:entropy-approx-02}) of the main text. 

\section{Relationship between $\Delta\epsilon_n$ and $\Delta\bar\epsilon_n$}
\label{sec:delta-eps-delta-eps-bar}
In this section, we derive Eq.~(\ref{eq:eps-eps-bar}) for a generic dispersion of the form $\epsilon_n=\varepsilon_0n^p$, for $\varepsilon\to0$, and $n=1,2,\dots$ and $p\geq 1$. First, we obtain the average dispersion
\begin{align}
\bar\epsilon_n&=\frac{\epsilon^\mathrm o_n+\epsilon^\mathrm e_n}{2}\\
    &=\frac{\varepsilon_0 }{2}\left[\left(2n+1\right)^{p}+\left(2n\right)^{p}\right].
\end{align}
Then, 
\begin{align}
    \Delta\bar{\epsilon}_{n}&=\bar{\epsilon}_{n+1}-\bar{\epsilon}_{n} \\
    &=\frac{\varepsilon_0 }{2}\left\{ \left(2n+3\right)^{p}+\left(2n+2\right)^{p}-\left[\left(2n+1\right)^{p}+\left(2n\right)^{p}\right]\right\}\\
    &=\frac{\varepsilon_0 }{2}\left[4\left(\begin{array}{c}
p\\
1
\end{array}\right)\left(2n\right)^{p-1}+12\left(\begin{array}{c}
p\\
2
\end{array}\right)\left(2n\right)^{p-2}+\dots\right],
\end{align}
where the line was obtained using a binomial expansion. Similarly, we find
\begin{align}
\Delta\epsilon_{n}&=\frac{\epsilon_{n}^{\text{o}}-\epsilon_{n}^{\text{e}}}{2} \\
    &=\frac{\varepsilon_0 }{2}\left[\left(2n+1\right)^{p}-\left(2n\right)^{p}\right]\\
    &=\frac{\varepsilon_0 }{2}\left[\left(\begin{array}{c}
p\\
1
\end{array}\right)\left(2n\right)^{p-1}+\left(\begin{array}{c}
p\\
2
\end{array}\right)\left(2n\right)^{p-2}+\dots\right].
\end{align}
Hence, 
\begin{align}
    \frac{\Delta\epsilon_{n}}{\Delta\bar{\epsilon}_{n}}&=\frac{\frac{\varepsilon_0 }{2}\left[\left(\begin{array}{c}
p\\
1
\end{array}\right)\left(2n\right)^{p-1}+\left(\begin{array}{c}
p\\
2
\end{array}\right)\left(2n\right)^{p-2}+\dots\right]}{\frac{\varepsilon_0 }{2}\left[4\left(\begin{array}{c}
p\\
1
\end{array}\right)\left(2n\right)^{p-1}+12\left(\begin{array}{c}
p\\
2
\end{array}\right)\left(2n\right)^{p-2}+\dots\right]}.
\end{align}
Consider the large-$n$ limit where
\begin{align}
    \lim_{n\to\infty}\frac{\Delta\epsilon_{n}}{\Delta\bar{\epsilon}_{n}}&=\lim_{n\to\infty}\frac{\left(\begin{array}{c}
p\\
1
\end{array}\right)+\left(\begin{array}{c}
p\\
2
\end{array}\right)\frac{\left(2n\right)^{p-2}}{\left(2n\right)^{p-1}}+\dots}{4\left(\begin{array}{c}
p\\
1
\end{array}\right)+12\left(\begin{array}{c}
p\\
2
\end{array}\right)\frac{\left(2n\right)^{p-2}}{\left(2n\right)^{p-1}}+\dots} \\
&= \frac{1}{4},
\end{align}
which is Eq.~(\ref{eq:eps-eps-bar}) of the main text.

\bibliography{apssamp}

\end{document}